\documentclass[twocolumn]{aastex63}

\usepackage[utf8]{inputenc}

\usepackage{hyperref}
\hypersetup{
    colorlinks=true,
    linkcolor=black,
    filecolor=magenta,      
    urlcolor=blue,
    }

\usepackage{graphicx}
\graphicspath{ {./figs/} }

\usepackage{multirow}

\usepackage{macros}
\input{macro_data_mbreak_is_gammalow}
\input{macro_data_A_0_mbreak_free}

\usepackage{amsmath}
\usepackage{amsfonts}
\usepackage{amssymb}

\begin{document}

\title{Bridging the Gap: Categorizing Gravitational-Wave Events at the Transition Between Neutron Stars and Black Holes \\
}

\author{Amanda Farah}
\email{afarah@uchicago.edu}
\affiliation{Department of Physics, University of Chicago, Chicago, IL 60637, USA}

\author{Maya Fishbach}
\altaffiliation{NASA Hubble Fellowship Program Einstein Postdoctoral Fellow}
\affiliation{Center for Interdisciplinary Exploration and Research in Astrophysics (CIERA) and Department of Physics and Astronomy,
Northwestern University, 1800 Sherman Ave, Evanston, IL 60201, USA}

\author{Reed Essick}
\affiliation{Perimeter Institute for Theoretical Physics, 31 Caroline Street North, Waterloo, Ontario, Canada, N2L 2Y5}

\author{Daniel E. Holz}
\affiliation{Department of Physics, University of Chicago, Chicago, IL 60637, USA}
\affiliation{Department of Astronomy and Astrophysics, Enrico Fermi Institute, and Kavli Institute for Cosmological Physics,\\University of Chicago, Chicago, IL 60637, USA}

\author{Shanika Galaudage}
\affiliation{School of Physics and Astronomy, Monash University, Clayton VIC 3800, Australia}
\affiliation{OzGrav: The ARC Centre of Excellence for Gravitational Wave Discovery, Clayton VIC 3800, Australia}

\date{\today}

\begin{abstract}
We search for features in the mass distribution of detected compact binary coalescences which signify the transition between neutron stars and black holes. We analyze all gravitational wave detections by LIGO-Virgo-KAGRA made through the end of the first half of the third observing run, and find clear evidence for two different populations of compact objects based solely on gravitational wave data. We confidently (99.4\%) find a deviation from a single power law describing neutron stars and low-mass black holes at $2.4^{+0.5}_{-0.5}M_{\odot}$, which is consistent with many predictions for the maximum neutron star mass. We find suggestions of the purported lower mass gap between the most massive neutron stars and the least massive black holes, but are unable to conclusively resolve it with current data. If it exists, we find the lower mass gap's edges to lie at $2.2^{+0.7}_{-0.5}M_{\odot}$ and $6.0^{+2.4}_{-1.4}M_{\odot}$. We re-examine events that have been deemed "exceptional" by the LIGO-Virgo-KAGRA collaborations in the context of these features. We analyze GW190814 self-consistently in the context of the full population of compact binaries, finding support for its secondary to be either a neutron star or a lower mass gap object, consistent with previous claims. Our models are the first to accommodate this event, which is an outlier with respect to the binary black hole population. We find that GW200105 and GW200115 probe the edges of, and may have components within, the lower mass gap. As future data improve global population models, the classification of these events will also become more precise.

\end{abstract}

\section{Introduction}
\label{sec:introduction}

The LIGO Scientific Collaboration, Virgo Collaboration, and KAGRA Collaboration (LVK) continue to expand the catalog of confidently detected Gravitational-Wave (GW) transients.
To date, there have been 46 unambiguous detections of binary black hole (BBH) mergers \citep{abbott_gwtc-1_2019,abbott_gwtc-2_2021}, two detections of binary neutron star (BNS) mergers \citep{abbott_gw170817_2017, abbott_gw190425_2020}, and, most recently, two neutron star-black hole (NSBH) merger candidates \citep{abbott_observation_2021}, not including the events reported in the most recent deep extended catalog \citep[GWTC-2.1]{abbott_gwtc-21_2021}.
However, some of the detected sources cannot obviously be ascribed to one of these source categories.
For example, GW190814 \citep{abbott_gw190814_2020} has a secondary mass of $m_2 = 2.59^{+0.08}_{-0.09}\, \Msun$, making it either the most massive neutron star (NS) or the lowest mass black hole (BH) detected to date.
GW190719\_215514, presented for the first time in GWTC-2.1 \citep{abbott_gwtc-21_2021}, is in a similar position.
These events lie at the edges of both the NSBH and BBH populations and therefore have the potential to probe the extremes of whichever subpopulation they belong.

However, their classification remains elusive, and in the absence of detectable tides or electromagnetic counterparts that definitively identify their secondaries as NSs, other classification schemes are necessary.
Some analyses have directly used NS equation of state (EOS) constraints to compute the probability that a given component mass is below the maximum allowed NS mass \citep{essick_discriminating_2020, abbott_gw190814_2020,abbott_observation_2021}.
In low latency, the LVK searches used a hard cutoff in component masses of $3\, \Msun$ in the third observing run\footnote{See the LIGO/Virgo Public Alerts User Guide: \url{https://emfollow.docs.ligo.org/userguide/content.html}} \citep{chatterjee_machine_2020} and $2.83\ \Msun$ in the second observing run \citep{abbott_low-latency_2019}, both motivated by EOS predictions for the maximum possible NS mass. 

However, modern models for the EOS predict different values for the maximum allowed neutron star mass, typically ranging anywhere between 2.0-2.5$\,\Msun$~\citep{margalit_constraining_2017, ruiz_gw170817_2018, shibata_constraint_2019,legred_impact_2021}.
See, e.g.,~\citet{chatziioannou_neutron-star_2020} for a review of recent observational constraints.
Additionally, the maximum mass of the galactic population of NSs is currently estimated to be $\sim 2-2.6\, \Msun$ \citep{antoniadis_millisecond_2016, alsing_evidence_2018, farr_population-informed_2020, fonseca_refined_2021} by electromagnetic observations of pulsars.

Meanwhile, electromagnetic observations of BHs in X-ray binaries suggest an absence of objects below $5 \Msun$ \citep{bailyn_mass_1998,ozel_black_2010, farr_mass_2011}.
This mismatch between the maximum observed NS mass and minimum observed BH mass points to the possibility of a ``lower mass gap'': a dearth of compact objects between $\sim 2.5$ and $\sim 5\, \Msun$.
GW detections have followed suit, with a lack of BBH detections in this range (see Fig.~\ref{fig:all-events}).

The low number of GW detections in the purported lower mass gap may be due to lower detector sensitivity to events with low masses (which are intrinsically quieter), but studies that take such selection effects into account find that the BBH distribution cannot be trivially extended to NS mass ranges.
\cite{fishbach_does_2020} found that a single power law cannot fit the mass distribution of both BBHs and BNSs via an analysis of events in GWTC-1 \citep{abbott_gwtc-1_2019}.
Similarly, \cite{abbott_population_2021} find a deviation from a power law in the BBH spectrum below $4\, \Msun$ using BBHs in GWTC-2.

We pursue a different approach, simultaneously modeling the overall distribution of all compact binaries without first dividing them into subpopulations.
In doing so, we may be less susceptible to issues arising from the placement of \textit{ad hoc} boundaries.
By examining the total mass distribution, we can first ask whether there is a need to subdivide the observed events into separate categories at all, such as events that fall above, below, or within the lower mass gap, instead of asserting such subpopulations exist \textit{a priori}.

A full description of the lower mass gap has been previously stymied by the small number of detections available.
Characterizing the nature of this feature in the overall mass spectrum of compact objects can reveal the true delineation between NSs and BHs in merging binaries, providing a useful compliment to event classification based on cutoffs that are either arbitrary, motivated by external observations of arguably different populations, or reliant on theoretical predictions.
With the recent influx of GW detections, enough data is available to make such classification schemes possible.

To resolve the transition between NSs and BHs, we follow the procedure of \cite{fishbach_does_2020} (herein FEH20) and model the full spectrum of compact binary coalesences (CBCs).
We do not separate events based on previous classifications, instead using all GW events available to determine if a global fit to component masses alone is able to find two or more distinct subpopulations.
Additionally, fitting a single mass model across the entire spectrum of detected CBCs allows for the inclusion of GW events that are not definitively classified into a specific source category.

We further describe our methodology and present the population models used in Sec.~\ref{sec:methods}.
We show that the transition between NSs and BHs cannot be described by a single power law and explore other morphologies  in Sec.~\ref{sec:results-1}.
In Sec.~\ref{sec:loo-0814}, we demonstrate that the models considered in this work are the first to be able to accommodate GW190814. We describe insights on the NSBH events in Sec.~\ref{sec:nsbh}.
We then classify all low-mass events with respect to the inferred subpopulation delineations in Sec.~\ref{sec:classification}, concluding in Sec.~\ref{sec:discussion}

\section{Methods}
\label{sec:methods}

We first describe our selection criteria for which events to include within our analysis in Sec.~\ref{sec:event list}.
We then describe our parametric models for the mass distribution in Sec.~\ref{sec:models} before describing metrics that can be used to describe features in the mass spectrum, sometimes without the need for a particular parametrization, in Sec.~\ref{sec:metrics of gap prominence}.
\subsection{Event List}
\label{sec:event list}

\begin{figure}
    \centering
    \includegraphics[width=1.0\columnwidth]{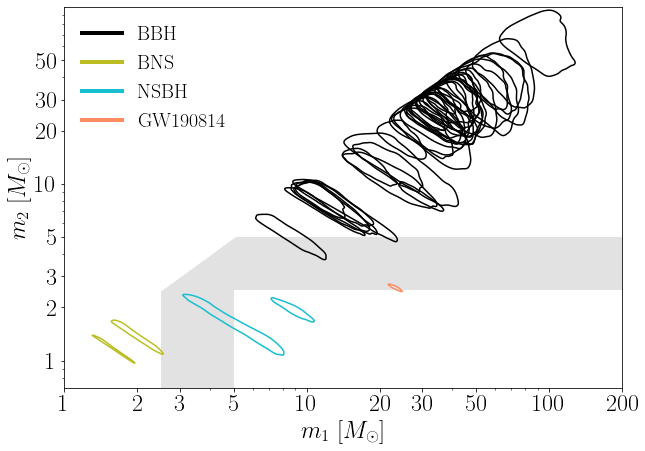}
    \caption{
    \label{fig:all-events}
        90\% posterior credible intervals for the component masses for all CBCs included in this population study assuming uniform priors in detector-frame masses.
        Events classified by the LVK as BBHs, BNSs and NSBHs are shown in black, green, and blue, respectively.
        The ambiguously classified event GW190814 is shown in orange.
        The grey band indicates the approximate location of the purported ``lower mass gap.''
        GW190814 is the only event within this region at more than 90\% credibility.
    }
\end{figure}

We consider all published CBC detections made by the LVK to date that have a matched-filter signal-to-noise ratio (SNR) greater than 11.
That is, all events in GWTC-2 \citep{abbott_gwtc-2_2021} that pass this threshold, as well as the two NSBH detections from the second half of the third observing run \citep[O3b]{abbott_observation_2021}).
None of the events newly presented in GWTC-2.1 pass this threshold \citep{abbott_gwtc-21_2021}.
This results in 46 events, all of which are shown in Fig.~\ref{fig:all-events}.
For the purposes of estimating sensitivity, we treat the NSBH events as if they were detected in the first half of the third observing run because data from other detections during O3b are not yet publicly available.
This choice introduces a modest bias in the inferred rate of NSBH-like events, which is discussed in Appendix~\ref{sec:loo_nsbh}

\subsection{Population Models}
\label{sec:models}

We consider several nested mass models.
That is, we describe our conclusions in the context of different sets of assumptions about the mass distribution, each of which corresponds to a specific hyperprior as described in Tab.~\ref{tab:hyperprior}.
Following \citet{fishbach_picky_2020}, FEH20, and \citet{doctor_black_2020} we parametrize the joint component mass distribution with separate draws for each component mass from a one-dimensional mass distribution and a pairing function.
Our one-dimension mass model can be described by a two-piece broken power law adorned with several multiplicative features that chip off parts of the mass distribution to produce dips and additional features.
That is, the one-dimensional distribution is described by
\begin{widetext}
\begin{equation}\label{eq:1D mass}
    p(m|\lambda_\textsc{1D}) \propto h(m|m_\mathrm{min}, \eta_\mathrm{min}) \times n(m|\gamma_\mathrm{low}, \gamma_\mathrm{high}, \eta_\mathrm{low}, \eta_\mathrm{high}, A) \times \ell(m|m_\mathrm{max}, \eta_\mathrm{max}) \left\{ \begin{matrix} \left(m/\mbreak\right)^{\alpha_1} & \text{if} & m < \mbreak \\ \left(m/\mbreak\right)^{\alpha_2} & \text{if} & \mbreak \leq m \end{matrix} \right.
\end{equation}
where
\begin{equation}
    \ell(m|m_0, \eta) = \left(1 + \left(\frac{m}{m_0}\right)^\eta\right)^{-1}
\end{equation}
is a low-pass Butterworth filter with roll-off mass $m_0$ and the sharpness of the roll-off set by $\eta$ (larger $\eta$ imply sharper roll-offs),
\begin{equation}
    h(m|m_0, \eta) = 1 - \ell(m|m_0, \eta) = \left(1 + \left(\frac{m_0}{m}\right)^\eta\right)^{-1}
\end{equation}
is a corresponding high-pass filter, and
\begin{equation}
    n(m|\gamma_\mathrm{low}, \gamma_\mathrm{high}, \eta_\mathrm{low}, \eta_\mathrm{high}, A) = 1 - A \times h(m|\gamma_\mathrm{low}, \eta_\mathrm{low}) \times \ell(m|\gamma_\mathrm{high}, \eta_\mathrm{high})
\end{equation}
\end{widetext}
is a notch filter that subtracts a fraction of the signal (set by $A$) between $\gamma_\mathrm{low}$ and $\gamma_\mathrm{high}$.
Our one-dimensional mass model builds upon a basic broken power law by adding a high-pass filter at the lowest masses to allow for a fixed but smooth turn-on of the mass distribution, a low-pass filter at the highest masses to allow for a variable tapering of the mass distribution, and a notch filter in between to model a potential lower mass gap.
This model has 12 parameters, which we collectively refer to as $\lambda_\textsc{1D}$. It was first presented in FEH20.
However, we include an additional high-pass filter at low masses instead of a sharp cut-off at $m_\mathrm{min}$.
FEH20's model is therefore nested in ours in the limit $\eta_\mathrm{min} \gg 1$.

To construct population models for both component masses, we employ a pairing function $f_p$ that relates primary and secondary masses:

\begin{multline}\label{eq:joint mass distrib}
    p(m_1, m_2| \lambda_\textsc{1D}, \beta_\mathrm{low}, \beta_\mathrm{high}) \propto \\ \Theta(m_2 \leq m_1) \times p(m_1 | \lambda_\textsc{1D}) \times p(m_2| \lambda_\textsc{1D}) \\ \times f_p(m_1,m_2|\beta_{\text{low}},\beta_{\text{high}}) \ ,
\end{multline}
where $\Theta(\cdot)$ is the Heaviside function that enforces the convention $m_2 \leq m_1$.

Note that the one-dimensional distribution's hyperparameters $\lambda_\textsc{1D}$ are shared between the primary and secondary mass distributions.
Note also that the one-dimensional distribution does not, in general, correspond to the marginal distribution for either the primary or secondary mass.
We use the pairing function
\begin{equation}\label{eq:pairing}
    f_p(m_1,m_2|\beta_{\text{low}},\beta_{\text{high}}) = \left\{ \begin{matrix}
    \left(m_2/m_1\right)^{\beta_{\text{low}}}  & \text{ if } & m_2 < 5\,\Msun\\
    \left(m_2/m_1\right)^{\beta_{\text{high}}} & \text{ if } & 5\,\Msun \leq m_2 \\
    \end{matrix} \right.
\end{equation}
to allow for the possibility that most BBHs have different formation channels than NS-containing or lower-mass-gap binaries.
We note that this is a different pairing function than that used in FEH20, which employed a single power law in mass ratio for all sources.
Again, their model is nested within ours in the limit $\beta_\mathrm{low} = \beta_\mathrm{high}$.

For simplicity, we assume fixed, independent spin and redshift distributions.
We assume sources are uniformly distributed in co-moving volume ($V_c$) and source frame time so that
\begin{equation}
    p(z) \propto \frac{dV_c}{dz} \left(\frac{1}{1+z}\right)
\end{equation}
We also assume spin distributions where each component is independent and follows a distribution that is uniform in magnitude $\chi_{1,2}$ and isotropic in orientation so that
\begin{align}
    p(|\chi|) &= \mathrm{U}(0, 1), \\
    p(z) &= \mathrm{U}(z)
\end{align}
where $\mathrm{U}(x, y)$ is the uniform distribution between $x$ and $y$ and $z_{1,2} = \cos{\theta_{1,2}}$ is the cosine of the tilt angle between component spin and a binary's angular momentum.

Although some authors have shown evidence for the evolution of the mass distribution and/or the merger rate with redshift \citep{fishbach_when_2021, abbott_population_2021}, we do not expect these choices to affect the mass inference within current statistical uncertainties.

\begin{figure}
    \centering
    \includegraphics[width=1.0\columnwidth]{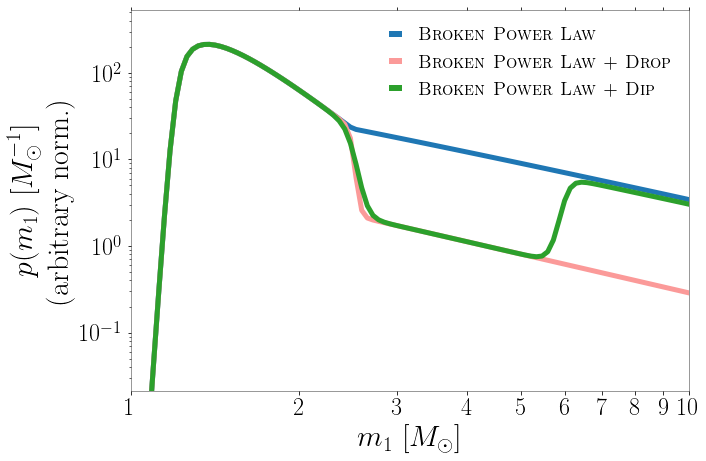}
    \caption{\label{fig:cartoons} Illustration of the primary mass distribution for the three mass models used in this work. All models are depicted with the same values for $\mbreak, \alpha_1,$ and  $\alpha_2$.}
\end{figure}

Eqs.~\ref{eq:1D mass} and~\ref{eq:pairing} intentionally provide a lot of flexibility in the functional form of the mass distribution via 14 free parameters while encoding several features predicted by theory.
Because the current data may not be able to constrain all the hyperparameters, we consider three nested models in order to demonstrate several of our conclusions more clearly.
In order of most to least model freedom,
\begin{itemize}
    \item \dipbreak{} (BPL+Dip): assumes the edges of low-mass features are steep but leaves the locations as free parameters, and fixes the break in the power law to occur at the lower edge of the gap. Sets $\eta_\mathrm{min} = \eta_\mathrm{high} = 50 \gg 1$ and assumes $\mbreak = \gamma_\mathrm{low}$. Alternatively, at times we assume the gap is deep with a sharp lower edge but allow the upper edge to be fit as a free parameter ($A=0.98$ and fits $\eta_\mathrm{high}$).
    \item \discontinuity{} (BPL+Drop): makes similar assumptions as BPL+Dip, except for $\eta_\mathrm{high} = 0$. This \textit{de facto} models a sharp drop-off above $\glo$ and removes the feature at $\ghi$. 
    \item \broken{} (BPL): assumes a steep turn-on at $m_\mathrm{min}$ but removes the dip/discontinuity by setting $\eta_\mathrm{low} = \eta_\mathrm{high} = 0$.
\end{itemize}
An illustration of these mass models is provided in Fig.~\ref{fig:cartoons}, and Tab.~\ref{tab:hyperprior} summarizes the hyperprior adopted for each of these models.

\begin{table*}
    \centering
    \begin{tabular}{p{1.50cm} p{1.25cm} c p{5.00cm} | c|c | c | c}
        \hline \hline
         \multicolumn{3}{c}{\multirow{3}{*}{\textbf{Parameter}}} & \multicolumn{1}{c}{\multirow{3}{*}{\textbf{Description}}} & \multicolumn{4}{c}{\textbf{Prior}} \\
          & & & & \multicolumn{2}{c|}{\multirow{2}{*}{\textbf{BPL+Dip}}} & \multirow{2}{*}{\textbf{BPL+Drop}} & \multirow{2}{*}{\textbf{BPL}} \\
        \hline \hline
         & \multirow{5}{*}{broken}
           & \multirow{2}{*}{$\mbreak$}       & dividing mass between high-mass and low-mass power laws & \multicolumn{4}{c}{\multirow{2}{*}{$m_\mathrm{break} = \gamma_\mathrm{low}$}} \\
        \cline{3-8}
         & \multirow{3}{*}{power law}
          & \multirow{2}{*}{$\alpha_1$}             & spectral index of power law below $\mbreak$ & \multicolumn{4}{c}{\multirow{2}{*}{$\mathrm{U}(-4, 12)$}} \\
        \cline{3-8}
         & & \multirow{2}{*}{$\alpha_2$}             & spectral index of power law above $\mbreak$ & \multicolumn{4}{c}{\multirow{2}{*}{$\mathrm{U}(-4, 12)$}} \\
        \cline{2-8}
         & & \multirow{2}{*}{$\gamma_\mathrm{low}$}  & roll-off mass for lower edge of the mass gap & \multicolumn{4}{c}{\multirow{2}{*}{$\mathrm{U}(1.4\, \Msun, 3.0\, \Msun)$}} \\
        \cline{3-8}
         & \multirow{3}{*}{notch}
          & \multirow{2}{*}{$\eta_\mathrm{low}$}    & sharpness of lower edge of the mass gap & \multicolumn{3}{c|}{\multirow{2}{*}{$\eta_\mathrm{low} = 50$}} & \multirow{2}{*}{N/A} \\
        \cline{3-8}
        \multirow{3}{*}{1D mass}
         & filter
           & \multirow{2}{*}{$\gamma_\mathrm{high}$} & roll-off mass for the upper edge of the mass gap & \multicolumn{2}{c|}{\multirow{2}{*}{$\mathrm{U}(3\,\Msun,9\,\Msun)$}} & \multicolumn{2}{c}{\multirow{2}{*}{N/A}} \\
        \cline{3-8}
        distribution
         & & \multirow{2}{*}{$\eta_\mathrm{high}$}   & sharpness of the upper edge of the mass gap &  \multirow{2}{*}{$\mathrm{U}(-4, 12)$} & \multirow{2}{*}{$\eta_\mathrm{high} = 50$} & \multirow{2}{*}{$\eta_\mathrm{high} = 0$} & \multirow{2}{*}{N/A} \\
        \cline{3-8}
         & & \multirow{2}{*}{$A$}                    & \multirow{2}{*}{depth of the notch filter} & \multirow{2}{*}{$A=0.98$} & \multirow{2}{*}{$\mathrm{U}(0, 1)$} & \multirow{2}{*}{$\mathrm{U}(0, 2)^{\dagger}$} & \multirow{2}{*}{$A=0$} \\
        \cline{2-8}
         & \multirow{3}{*}{high-pass} & \multirow{2}{*}{$m_\mathrm{min}$}       & roll-off mass for high-pass filter at the lowest masses allowed & \multicolumn{4}{c}{\multirow{2}{*}{$\mathrm{U}(1.0\,\Msun,1.4\,\Msun)$}} \\
        \cline{3-8}
         & \multirow{1}{*}{filter}
          & \multirow{2}{*}{$\eta_\mathrm{min}$}    & sharpness of high-pass filter at the lowest masses allowed & \multicolumn{4}{c}{\multirow{2}{*}{$\eta_\mathrm{min}=50$}} \\
        \cline{2-8}
        
         & \multirow{3}{*}{low-pass} & \multirow{2}{*}{$m_\mathrm{max}$}       & roll-off mass for low-pass filter at the highest masses allowed & \multicolumn{4}{c}{\multirow{2}{*}{$\mathrm{U}(30\,\Msun, 100\,\Msun)$}} \\
        \cline{3-8}
         & \multirow{1}{*}{filter}
          & \multirow{2}{*}{$\eta_\mathrm{max}$}    & sharpness of low-pass filter at the highest masses allowed & \multicolumn{4}{c}{\multirow{2}{*}{$\mathrm{U}(-4, 12)$}} \\
        \hline
        \multirow{3}{*}{pairing}
         & \multicolumn{2}{c}{\multirow{2}{*}{$\beta_1$}} & spectral index of pairing function if $m_2 < 5\,\Msun$ & \multicolumn{4}{c}{\multirow{2}{*}{$\mathrm{U}(-4,12)$}} \\
        \cline{2-8}
        function
         & \multicolumn{2}{c}{\multirow{2}{*}{$\beta_2$}} & spectral index of pairing function if $5\,\Msun \leq m_2$ & \multicolumn{4}{c}{\multirow{2}{*}{$\mathrm{U}(-4,12)$}} \\
        \hline
    \end{tabular}
    \caption{
    Hyperparameters of our mass model and hyperpriors corresponding to specific nested models: \dipbreak{} (BPL+Dip), \discontinuity{} (BPL+Drop), and \broken{} (BPL).
    We denote the uniform distribution between $x$ and $y$ as $\mathrm{U}(x, y)$, list specific values that are fixed in some priors, and denote when hyperparameters are irrelevant to a specific nested model with ``N/A''.
    Note that there are two priors that we refer to as BPL+Dip since they explore similar phenomenology: one assumes a sharp upper edge of the gap but variable gap depth and the other assumes a deep gap but allows for smooth edges.
    We specify which priors were used in context within the text. $^{\dagger}${The allowed values of $A$ within \dipbreak{} are between $0$ and $1$, whereas the elimination of the upper edge of the notch filter in \discontinuity{} allows the values of $A$ to range between $0$ and $2$ in that case.}
    }
    \label{tab:hyperprior}
\end{table*}

\subsection{Characterizing the Prominence of Features in the Mass Distribution}
\label{sec:metrics of gap prominence}

In the case of a mass gap with sharp edges ($\eta_{\mathrm{low},\mathrm{high}} \gg 1$), the notch filter strength $A$ is an unambiguous measure of the gap depth, and the parameters $\glo$ and $\ghi$ are the precise locations of the beginning and end of the gap, respectively.
With this in mind, we can quantify the evidence in favor of a gap with a Bayes factor between models where $A\neq0$ and $A=0$: $\mathcal{B}^{A\neq0}_{A=0}$.

However, this assumes the shape of the gap is well known \textit{a priori}.
This is not the case, particularly for the upper edge of the gap.
In the case of smooth gap edges, the depth and location of the gap are dictated by a more complex combination of hyperparameters.
We therefore define a few straightforward measures of gap prominence which can be used regardless of the exact parametrization adopted.

The first is the existence of a local minimum in the mass distribution between $1$\mdash{}$10 \Msun$.
A lack of support for such a dip in the mass distribution in this region would indicate either a lack of a lower mass gap or an inability to resolve one if it does exist.
Throughout this text, we report Bayes factors for the existence of a local minimum that are defined as the ratio of the percentage of draws from the hyperposterior that find a local minimum to the percentage of draws from the hyperprior that find a local minimum.

The second is the relative probability of events between $\glo$ and $\ghi$ vs. that between $m_{\rm min}$ and $\glo$:
\begin{equation}
    r^{\rm NS}_{\rm gap} = \frac{p(m\in[m_\mathrm{min}, \glo)\,|\,\lambda_\textsc{1D})}{p(m\in[\glo, \ghi)\,|\,\lambda_\textsc{1D})}.
    \label{eq:rel-rates-ns}
\end{equation}
With this definition, $r^{\rm NS}_{\rm gap} \gg 1$ indicates a prominent drop-off in the mass distribution at $\glo$, whereas $r \sim 1$ is indicative of a smoother transition above NS masses.

The third is similar to the second: it is the relative probability of events between $\glo$ and $\ghi$ vs. that in an equally wide mass interval starting at $\ghi$:
\begin{equation}
    r^{\rm BH}_{\rm gap} = \frac{p(m\in[\ghi, 2\ghi - \glo)\,|\,\lambda_\textsc{1D})}{p(m\in[\glo, \ghi)\,|\,\lambda_\textsc{1D})}.
    \label{eq:rel-rates-bh}
\end{equation}
A value of $r^{\rm BH}_{\rm gap} \gg 1$ indicates a very prominent lower mass gap, whereas $r$ consistent with unity indicates a flat transition between NS and BH masses.

\subsection{Statistical Framework}
\label{sec:statistical framework}

Using the parametrized distributions from Sec.~\ref{sec:models},
we construct a hierarchical Bayesian inference to determine the appropriate population-level parameters, $\Lambda = \{\lambda_\textsc{1D}, \beta_\mathrm{low}, \beta_\mathrm{high}\}$ given the observed set of data $\{D_j\}$ for $N$ observed events \citep[see, e.g.,][ for more details]{loredo_accounting_2004, thrane_introduction_2019, mandel_extracting_2019}.
We model the data as an inhomogeneous Poisson process with the rate density (expected number of events per unit time per single-event-parameter hypervolume) given by
\begin{equation}
    \frac{d\mathcal{N}}{dm_1 dm_2, ds_1 ds_2 dz} = \mathcal{R} p(z) p(s_1) p(s_2) p(m_1,m_2|\Lambda)
\end{equation}
where $\mathcal{R}$ acts as a normalizing constant that sets the overall magnitude of the rate.
The posterior on the population hyper-parameters, assuming a prior on the overall rate of mergers $p(\mathcal{R}) \sim 1/\mathcal{R}$ and marginalizing, is
\begin{equation}
    p(\Lambda |\{D_j\}) = p(\Lambda) \prod_j^N \frac{p(D_j|\Lambda)}{\mathcal{E}(\Lambda)},
    \label{eq:hyperposterior}
\end{equation}
where
\begin{widetext}
\begin{equation}\label{eq:marginal likelihood}
    p(D_j|\Lambda) = \int dm_1 dm_2 ds_1 ds_2 dz \, p(z) p(s_1) p(s_2) p(m_1, m_2|\Lambda) p(D_j|m_1, m_2, s_1, s_2, z)
\end{equation}
is the marginal likelihood for the $j^{\rm th}$ event,
\begin{equation}\label{eq:selection effects}
    \mathcal{E}(\Lambda) = \int dm_1 dm_2 ds_1 ds_2 dz \, p(z) p(s_1) p(s_2) p(m_1, m_2|\Lambda) P(\mathrm{det}|m_1, m_2, s_1, s_2, z)
\end{equation}
\end{widetext}
is the fraction of detectable events in a population described by $\Lambda$, and $P(\mathrm{det}|m_1, m_2, s_1, s_2, z)$ is the probability that any individual event with parameters $m_1$, $m_2$, $s_1$, $s_2$, and $z$ would be detected, averaged over the duration of the experiment.



If we wish to simultaneously infer the properties of individual events along with the population hyperparameters, we simply do not marginalize over those events' parameters and obtain, e.g.,
\begin{multline}
    p(m_1^{(i)}, m_2^{(i)}, \Lambda | \{D_j\} ) = \\
    p(D_i|m_1^{(i)}, m_2^{(i)}) p(m_1^{(i)}, m_2^{(i)} | \Lambda) \\
    \times \frac{p(\Lambda)}{\mathcal{E}(\Lambda)} \prod_{j\neq i}^N \frac{p(D_j|\Lambda)}{\mathcal{E}(\Lambda)}
\end{multline}
where $p(D_i|m_1^{(i)}, m_2^{(i)})$ is still marginalized over $s_1^{(i)}$, $s_2^{(i)}$, and $z^{(i)}$.
We typically investigate the different levels of our hierarchical inference separately, though.
That is, we examine the hyperposterior for $\Lambda$ in the population-level inference after marginalizing over uncertainty in individual-event parameters, or we examine the event-level population-informed posteriors for individual event parameters after marginalizing over the uncertainty in the inferred population.

In practice, the high-dimensional integrals in Eqs.~\ref{eq:marginal likelihood} and~\ref{eq:selection effects} are approximated via importance sampling.
That is, given a set of $N_j$ event-level posterior samples for the $j^\mathrm{th}$ event drawn with a reference prior $p_\mathrm{ref}(m_1, m_2, s_1, s_2, z)$, we approximate
\begin{equation}\label{eq:approx marginal likelihood}
    p(D_j|\Lambda) \approx \frac{1}{N_j}\sum\limits_\alpha^{N_j} \frac{p(m_1^{(\alpha)}, m_2^{(\alpha)}, s_1^{(\alpha)}, s_2^{(\alpha)}, z^{(\alpha)}|\Lambda)}{p_\mathrm{ref}(m_1^{(\alpha)}, m_2^{(\alpha)}, s_1^{(\alpha)}, s_2^{(\alpha)}, z^{(\alpha)})}
\end{equation}
Similarly, by simulating a large set of $M$ signals drawn from an injected population $p_\mathrm{inj}$, we can approximate Eq.~\ref{eq:selection effects} with a sum over the subset of $m$ detected signals
\begin{equation}\label{eq:approx selection effects}
    \mathcal{E}(\Lambda) \approx \frac{1}{M}\sum\limits_\alpha^m \frac{p(m_1^{(\alpha)}, m_2^{(\alpha)}, s_1^{(\alpha)}, s_2^{(\alpha)}, z^{(\alpha)}|\Lambda)}{p_\mathrm{inj}(m_1^{(\alpha)}, m_2^{(\alpha)}, s_1^{(\alpha)}, s_2^{(\alpha)}, z^{(\alpha)})}
\end{equation}
We approximate the detectable set of signals as those that yield an optimal network signal-to-noise ratio (SNR) $\geq 11$ based on networks of the LIGO Hanford and Livingston detectors during O1 and O2 as well as the LIGOs and Virgo during O3a.
During each observing run, we use a single representative power spectral density (PSD) to model each detector's sensitivity separately \citep{lvk_l1_2015,lvk_h1_2015,lvk_l1_2017,lvk_h1_2017,lvk_aligo_2020} and estimate the network SNR by coherently projecting simulated signals into all detectors.
While this semi-analytic approach to sensitivity estimation is commonplace in the literature, we note that it makes several assumptions that may not be perfectly accurate (a single stationary PSD for each detector, the same SNR threshold for all systems, etc.).
The size of the possible systematic errors these assumptions could introduce is not known precisely, but we believe it is less than $\mathcal{O}(10\%)$ based on comparisons of our semianalytic rate estimates with fixed populations to those previously published in the literature using real searches \citep{abbott_gwtc-1_2019, abbott_gwtc-2_2021}.

We sample from the posterior distribution in Eq.~\ref{eq:hyperposterior} using the approximations in Eqs.~\ref{eq:approx marginal likelihood} and~\ref{eq:approx selection effects} to determine the shape of the mass distribution using \texttt{gwpopulation} \citep{talbot_parallelized_2019}. 
Furthermore, where needed, we estimate Bayes factors via Savage-Dickey Density Ratios \citep{dickey_weighted_1970,wagenmakers_bayesian_2010} using the hyperposteriors and the hyperpriors described in Tab.~\ref{tab:hyperprior}.

\section{Population-Level Insights}
\label{sec:results-1}

In this section, we search for and characterize features between NS-like and BH-like masses.
Following FEH20, we first determine the existence of such a feature by asking whether a single power law can describe the CBC mass spectrum between $1 - 10 \Msun$ in Sec.~\ref{sec:broken}.
We find that it cannot. We then describe the nature of this deviation from a power law in Sec.~\ref{sec:dipbreak} through the features provided by the parametrized models from Sec.~\ref{sec:models}.

\subsection{Deviation from a Single Power Law Describing BH and NS Masses}
\label{sec:broken}

Neither FEH20 nor \cite{abbott_population_2021} include GW190814, nor do they include GW190425, a BNS detected in the first half of the third LVK observing run \citep{abbott_gw190425_2020}.
NSBH-like events GW200105\_162426 and GW200115\_042309 (abbreviated GW200105 and GW200115), which contain secondaries near the upper limits of what can be considered a NS and primaries at lower BH masses than have been previously detected in definitive BBHs, had not been published at the time of either study.
These events appear to bridge the NS and BH subpopulations, so the deviations from a single power law in the range $1-10\,\Msun$ may no longer be present in the most recent data. We therefore aim to determine if the inclusion of these events removes the feature found by FEH20 and \cite{abbott_population_2021}.

\begin{figure}
    \centering
    \includegraphics[width=1.0\columnwidth]{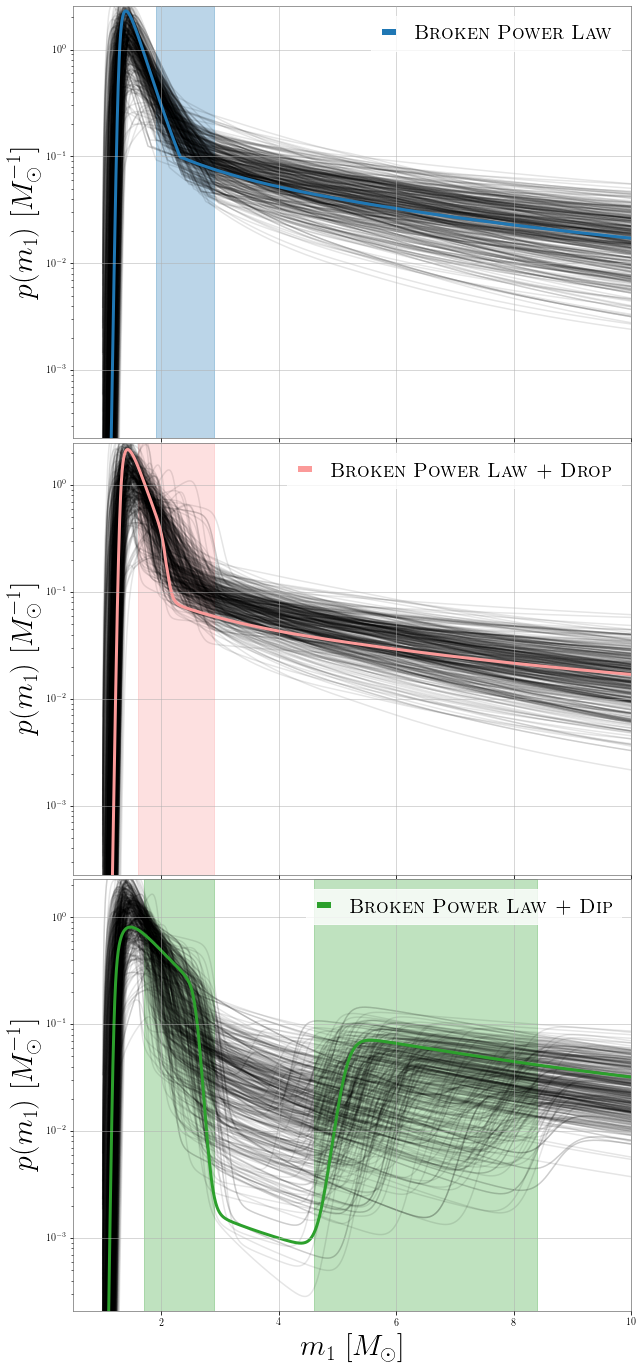}
    \caption{
        \label{fig:trace_plots} Draws from the inferred mass distributions between $1-10 \Msun$ using the \broken{} (\emph{top}), \discontinuity{} (\emph{middle}), and \dipbreak{} (\emph{bottom}) models.
        Each panel has 200 random draws from the hyperposterior (\emph{thin black lines}).
        As an illustration, the draw that maximizes the posterior is highlighted in color.
        90\% credible intervals on the locations of features in the mass distribution are indicated by colored bands.
        All three models find a steep falloff in the mass distribution between $\sim 2.2 - 3 \Msun$ and \dipbreak{} finds a subsequent rise at $\sim 6 \Msun$.
    }
\end{figure}

The first mass model we fit is \broken.
It has one power law spectral index $\alpha_1$ at NS-like masses and another $\alpha_2$ at BH-like masses, with a transition between the two at $\mbreak$.
Draws from the hyperposterior for \broken{} are shown in the top panel of Fig.~\ref{fig:trace_plots}. 
There is a clear transition in the mass distribution at $m_{\mathrm{break}} = \CIPlusMinus{\brokenMatterMattersBinnedPairing[param][NSmax]}\, \Msun$.
This value is consistent with current constraints on the maximum mass of cold, non-spinning NS, which are typically between 2.0-2.5$\,\Msun$ at 90\% confidence~\citep[see, e.g.,][]{essick_direct_2020, legred_impact_2021, farr_population-informed_2020, chatziioannou_neutron-star_2020}, although larger values are still consistent with low-density nuclear physics.

We find $\alpha_2 -\alpha_1 > 0$ at \aoneisnotatwopercent\% credibility, indicating that a single power law is unable to fit the mass distribution down to low masses. 
This is consistent with findings in FEH20 and \cite{abbott_population_2021}, and therefore the low mass events GW190814, GW200105, and GW200115 do not fully bridge the BH and NS populations.

As an aside, we also note that $\beta_2-\beta_1 > 0$ at \boneisnotbtwopercent\% credibility, indicating that the pairing functions between BBHs and CBCs with low secondary mass are distinct.
However, because of the fixed location of the break between $\beta_1$ and $\beta_2$, we do not use these parameters to locate delineations between subpopulations.

\subsection{Steep Decline After NS Masses}
\label{sec:discontinuity}
The steepness of the NS part of the power law in \broken{} (i.e. large magnitude of $\alpha_1$) may be driven primarily by a difference in the merger rates of BNS and BBH systems, rather than by the shape of the NS distribution itself.
While our particular analysis has a slight bias in the relative heights of the low mass and high mass parts of the mass distribution because of the choice to treat the NSBH events as detected in the second half of the third observing run, many other studies have found the BNS merger rate to be several times higher than the BBH merger rate \citep{abbott_gwtc-1_2019,abbott_binary_2019,fishbach_does_2020,abbott_gwtc-2_2021, abbott_population_2021}.

To decouple the shape of the NS distribution from the difference in merger rates between BNS and BBH systems, we we turn to \discontinuity.
This model allows for a drop-off in the mass distribution at $\mbreak$. 
The extent of the discontinuity is parameterized by $A$, which is a free parameter.
$A=2$ corresponds to a maximal discontinuity: in this case, the rate of BH-containing events drops to zero after $\mbreak$.
$A=0$ corresponds to no discontinuity \mdash{} in this case, \discontinuity{} reduces to \broken.
The resulting fit to this model is shown in the middle panel of Fig.~\ref{fig:trace_plots}.
A clear drop-off is found after NS masses, which can be quantified by $r^{\rm NS}_{\rm gap}$, the ratio of probability mass below $\mbreak$ to that in a  comparable interval just above it (Eq.~\ref{eq:rel-rates-ns}).
We find $r^{\rm NS}_{\rm gap} = 4.1^{+10}_{-3.0}$, which excludes 1 to $>90\%$ credibility.

\begin{figure}
    \centering
    \includegraphics[width=1.0\columnwidth]{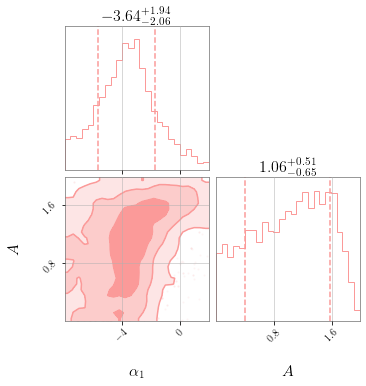}
    \caption{Values of the spectral index governing the low mass part of the spectrum $\alpha_1$ and drop depth $A$ inferred by the \discontinuity{} model.
    Contours indicate 1-$\sigma$, 2-$\sigma$, and 3-$\sigma$ credible regions.
    The two parameters are correlated and the lower right region is excluded: mild power law slopes are only allowed when $A$ is large.
    Thus, the inference finds a dramatic drop-off in the merger rate, either through a steep negative initial power-law slope ($\alpha_1$), or a discontinuity (large $A$).
    The two scenarios are both allowed by the data due to the inability to resolve the shape of the NS distribution.
    \label{fig:eta_2_0_A_vs_a1}
    }
\end{figure}

Fig.~\ref{fig:eta_2_0_A_vs_a1} shows the inferred posteriors for $A$ and $\alpha_1$ under the \discontinuity{} framework. 
These confirm the need for a sharp drop-off in the mass distribution at $\mbreak$, either by a large negative $\alpha_1$ or by a large $A$.
However, at the largest values of $A$, $\alpha_1$ is unconstrained.
We therefore conclude that the steep drop-off in the merger rate after NS masses is driving the constraints on $\alpha_1$, rather than the shape of the NS distribution itself. 
This is consistent with the findings of \cite{landry_mass_2021}, who show that the mass distribution of NSs in binaries is not yet able to be determined from GW data alone.

Within \discontinuity, the location of the break in the power law (and therefore the location of the discontinuity) is inferred to be at $\mbreak=2.20^{+0.51}_{-0.41} \Msun$, which is consistent with but shifted lower than the corresponding value inferred by \broken{} ($\CIPlusMinus{\brokenMatterMattersBinnedPairing[param][NSmax]}$).
This is due to the different modelling assumptions inherent in both models, and both values are equally plausible.
Furthermore, given the one-dimensional posterior on $A$, we find that the data cannot distinguish between $A=0$ and $A\neq0$ and therefore have no clear preference for \broken{} or \discontinuity.

\subsection{Searching for a Lower Mass Gap}
\label{sec:dipbreak}

Having confidently established a deviation from a single power law in the form of a steep decrease in the mass distribution after NS-like masses, we wish to identify other features.
In particular, we search for signatures of the purported lower mass gap. 

To do this, we employ the \dipbreak{} model, described in Sec.~\ref{sec:models}.
The fit to this model is shown in the bottom panel of Fig.~\ref{fig:trace_plots} assuming $\eta_\mathrm{high}=50$. 
We infer the depth of a potential mass gap through the hyperparameter $A$, where $A=0$ reduces to the \broken{} mass model and $A=1$ corresponds to a maximal gap.
The lower and upper edges of this gap are represented by $\glo$ and $\ghi$, respectively.
We infer $\glo =\CIPlusMinus{ \dipbreakMatterMattersBinnedPairing[param][NSmax]} \Msun$ and $\ghi~=~\CIPlusMinus{\dipbreakMatterMattersBinnedPairing[param][BHmin]} \Msun$.
These are consistent with the values found in FEH20 ($\glo = 2.2^{+0.6}_{-0.5} \Msun$, $\ghi = 6.7^{+1.0}_{-1.5} \Msun$).
This agreement with a GWTC-1 analysis suggests that the addition of events in GWTC-2 and the NSBH-like events are consistent with this previously inferred population and do not fully fill the lower mass gap, if one exists.
These events may, however, indicate that a lower mass gap might not be completely empty.

\begin{figure}
    \centering
    \includegraphics[width=1.0\columnwidth]{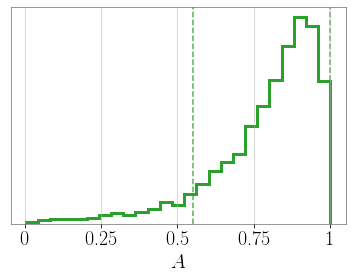}
    \caption{Posterior on the gap depth parameter $A$ under the \dipbreak{} model including all events.
        $A=1$ indicates a maximal gap, and $A=0$ indicates no mass gap.
        $A$ peaks at 0.90, with 90\% highest density interval between 0.55 and 1.0 (\emph{vertical dashed lines}), indicating support for both a partially-filled and totally empty mass gap.
    }
    \label{fig:A_mbreak_is_glo}
\end{figure}

Within the framework of \dipbreak, we find suggestions of a lower mass gap, but are unable to conclusively detect or rule out its existence.
The posterior on $A$ is shown in Fig.~\ref{fig:A_mbreak_is_glo}.
It peaks at large values but does not rule out the possibility that some events occupy the gap.
A model with a maximal gap ($A=1$) is favored over no gap ($A=0$) with a Bayes factor of \OneVsZeroBF, and $r^{\rm BH}_{\rm gap} = 1.84^{+5.74}_{-1.14}$, which is suggestive of the existence of a lower mass gap ($r^{\rm BH}_{\rm gap}$ is defined in Eq.~\ref{eq:rel-rates-bh}).

However, we find a Bayes factor of only 1.1 in favor of a local minimum between $1-10\, \Msun$, indicating that the assumption of sharp gap edges may be driving the other metrics of gap prominence\footnote{\localminimadipbreakpercent\% of hyperposterior samples produce a local minimum between $1\Msun$ and $10\Msun$, but this must be compared with the percentage of prior draws that find the same feature, which is \localminimadipbreakpriorpercent\%.}.

Due to the symmetric nature of the notch filter $n(m| A, \glo, \ghi, \eta_{\text{low}}, \eta_{\text{high}})$ used in \dipbreak{}, a sharp drop-off in the mass distribution after $\glo$ will always be accompanied by a subsequent rise at $\ghi$ when we set $\eta_{\mathrm{low}} = \eta_{\mathrm{high}\}} \gg 1$.
Thus, if the data prefer such a sharp drop-off that is not achieved through a steep $\alpha_1$, using \dipbreak{} may artificially inflate our metrics of gap significance. 
This systematic is due in part to the choice to set $\eta_{\text{low}} = \eta_{\text{high}} = 50$, rather than leaving the steepness of the gap edges as free parameters.
We therefore seek to determine if the data actually prefer a subsequent rise in the mass distribution after the initial drop at $\glo$ by allowing $\eta_{\text{high}}$ to be a free parameter. 
Positive values of $\eta_{\text{high}}$ indicate the existence of a subsequent rise, and larger positive values imply steeper rises.
To lessen the number of free parameters being fit to a region with relatively few detections, we enforce a drop in the mass distribution after NS masses by fixing $\eta_{\text{low}} = 50$ and $A=0.98$.
We are therefore asking, given the existence of a drop in the mass distribution at $\glo$, do the data necessitate a subsequent rise? 

\begin{figure}
    \centering
    \includegraphics[width=1.0\columnwidth]{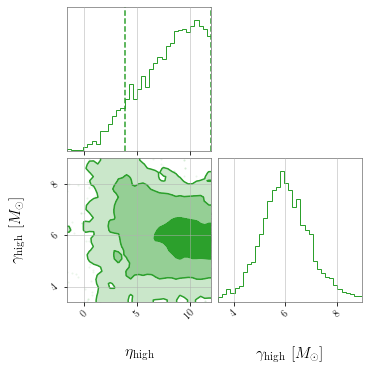}
    \caption{Corner plot of the parameters governing the location ($\ghi$) and steepness ($\eta_{\rm high}$) of the upper edge of the mass gap in the case where a steep decline in the mass distribution is enforced after NS masses ($A=0.98$).
    The 90\% highest probability density interval for $\eta_{\rm high}$ is indicated by vertical dashed lines.
    $\eta_{\rm high}$ is found to be positive, indicating a rise in the mass distribution near $\ghi$ after the drop enforced at $\glo$.
    }
    \label{fig:A_1_eta2_vs_ghi}
\end{figure}

Fig.~\ref{fig:A_1_eta2_vs_ghi} shows the two dimensional posteriors on $\ghi$ and $\eta_{\text{high}}$, the parameters determining the location and steepness of the upper edge of the mass gap.
We find $\eta_{\text{high}} >0$ at $99.8\%$ credibility, with lower 90\% credible bound at  $\eta_{\text{high}} = 2.82$, indicating that if there is a large drop in the mass distribution at $\glo$, the data prefer a subsequent rise at $\ghi$ rather than an additional drop ($\eta_{\text{high}} <0$) or a featureless transition ($\eta_{\text{high}} = 0$).
We additionally find a Bayes factor of 1.14 in favor of a local minimum between $1-10\, \Msun$.

In this model fit, $\ghi = 5.98^{+0.99}_{-0.87}\, \Msun$, which is completely consistent with the location of the global maximum of the BBH distribution found in \cite{abbott_population_2021}.
They find $7.8^{+1.8}_{-2.0}\, \Msun$ for the \textsc{Power Law + Peak} model and $6.02^{+0.78}_{-1.96}\, \Msun$ for the \textsc{Broken Power Law} model, the latter of which is more morphologically similar to the high-mass behavior of the models considered in this work.

As a final test of our results, we perform a fit where all hyperparameters are allowed to vary.
The results are described in Appendix~\ref{sec:all_params_free}, and we find a Bayes factor of $1.63$ in favor of a local minimum.
This indicates slightly more evidence for a lower mass gap, but is far from conclusive.
However, due to the small number of detections at low masses, we cannot meaningfully constrain more than two hyperparameters in this region.
We therefore fix $\eta_{\rm high} = 50$ and allow $A$ to vary for the remaining analyses, noting that a steep upper edge and variable gap depth allows for more intuitive interpretations of the event classification presented in Sec.~\ref{sec:classification}.

\section{Event-Level Insights}
\label{sec:methods-2}

Building upon our population-level insights, we now place individual events in the context of the inferred features in the mass distribution.
Sec.~\ref{sec:classification} discusses how we do this in general, marginalizing over both the uncertainty in individual events' parameters as well as the population hyperparameters.
We then discuss the implications for specific events that previous models have not been able to accommodate: GW190814 (Sec.~\ref{sec:loo-0814}) and the two NSBH-like events GW200105 and GW200115 (Sec.~\ref{sec:nsbh}).

\subsection{Event Classification}
\label{sec:classification}
Fits to the full spectrum of CBCs allow us to classify events with respect to  features inferred from the detected population.

Tab.~\ref{tab:evs-broken} uses the \broken{} model fit to characterize events in relation to $\mbreak$.
It lists the probabilities that each event's component masses lie in regions of interest.
For example, if one interprets $\mbreak$ to be the maximum NS mass, the probability that an event contains a NS is $P( m_2 \leq \mbreak)$ and the probability that an event is a BNS is $P( m_1 \leq \mbreak)$,  since $m_2 \leq m_1$ by definition.
Similarly, if we consider objects with masses larger than $\mbreak$ to be BHs, the probability that an event is a NSBH is $P(m_2 \leq \mbreak < m_1)$.
Using these assumptions, we find that all probabilities presented in Tab.~\ref{tab:evs-broken} are qualitatively consistent with the classifications given to each event by the LVK.

GW170817 is found here to have $P(m_1 \leq \mbreak) = \brokenMatterMattersBinnedPairing[prob_in_gap][GW170817][prob_mass1_NS]$, which is consistent with the LVK's classification of this event as a BNS. 
GW190425 has $P(m_1 \leq \mbreak) =  \brokenMatterMattersBinnedPairing[prob_in_gap][S190425z][prob_mass1_NS]$, with $< 10\%$ support for a NSBH classification.
This is consistent with this event's classification as a BNS.
NSBH events GW200105 and GW200115 both 
have larger $P(m_2 \leq \mbreak < m_1)$ than their respective $P(m_1 < \mbreak)$ or $P(m_2 > \mbreak)$, so under this framework they are more likely NSBHs than BNSs or BBHs.
The secondary mass of GW190814 is neither definitively a NS nor a BH under this classification scheme: the posterior on $\mbreak$ overlaps almost completely with that of GW190814's $m_2$.
%

Tab.~\ref{tab:evs-dipbreak} uses the \dipbreak{} model fit to characterize events in relation to $\glo$ and $\ghi$.
Events that have been previously identified by the LVK as containing a NS 
(GW170817, GW190425, GW200105, GW200115) all have $P(m_2 \leq \glo) > 0.95$.
BNS event GW170817 has $P(m_1 \leq \glo) >0.95$ as well, but BNS event GW190425 has $\sim10\%$ support of its primary being in the mass gap rather than a NS.
This is likely due to the relatively large uncertainties on both the component masses of GW and on $\glo$.
Events previously identified by the LVK as being consistent with NSBHs \citep[GW200105, GW200115]{abbott_observation_2021} receive a similar classification in this analysis, with large $P(m_2 \leq \glo)$, and near-vanishing $P(m_1 \leq \glo)$.

Additionally, the component masses of three events have considerable support in the mass gap: the secondary of GW190814, the secondary of GW190924\_021846, and the primary of GW200115. 
While a totally empty gap is not ruled out \mdash{} none of the three have more than 75\% support in the gap \mdash{} the combined effect of these three events is to decrease the inferred depth of the lower mass gap.
The secondary mass of GW190814 remains ambiguously classified: it has non-negligible support of being either in the lower mass gap or being a NS.
As the posterior on $m_2$ for GW190814 is well constrained, the ambiguous classification is instead due to the inference's inability to determine whether the event should be positioned in the first power law or in the notch filer of \dipbreak.
This results in bimodality in $\glo$'s posterior distribution, which we discuss further in Sec.~\ref{sec:loo-0814}.
As more detections of GW events constrain the global mass distribution, uncertainties in the locations of $\glo$ and $\ghi$ will decrease, allowing for more a more definitive classification of this event.

The probabilities in Tabs.~\ref{tab:evs-broken} and \ref{tab:evs-dipbreak} are marginalized over the uncertainties in each event's component masses and the uncertainties in the relevant model hyperparameters.

We further examine three events that were published as ``exceptional'' by the LVK. 

\begin{table*}
  \centering
  \begin{tabular}{c  c  c  c }
          \hline \hline
          \textbf{Probability}
                & $P( m_2 \leq \mbreak)$ 
                & $P( m_1 \leq \mbreak)$
                & $P(m_2 \leq \mbreak < m_1)$ \\
          \hline
          \textbf{Interpretation}
                & $P(\text{contains a NS})$
                & $P(\text{is a BNS})$
                & $P(\text{is a NSBH})$ \\
          \hline \hline
          GW170817
                & $>0.99$
                & \brokenMatterMattersBinnedPairing[prob_in_gap][GW170817][prob_mass1_NS]
                & \brokenMatterMattersBinnedPairing[prob_in_gap][GW170817][prob_NSBH]

                \\
          \hline
          GW190425
                & $>0.99$
                & 0.90
                & \brokenMatterMattersBinnedPairing[prob_in_gap][S190425z][prob_NSBH]
                \\
          \hline
          GW190814
                & \brokenMatterMattersBinnedPairing[prob_in_gap][S190814bv][prob_mass2_NS]
                & $<0.001$
                & \brokenMatterMattersBinnedPairing[prob_in_gap][S190814bv][prob_NSBH]
                \\
                \hline
          GW200105
                & \brokenMatterMattersBinnedPairing[prob_in_gap][GW200105][prob_mass2_NS]
                & $<0.001$
                & \brokenMatterMattersBinnedPairing[prob_in_gap][GW200105][prob_NSBH]
                \\
                \hline
          GW200115
                & \brokenMatterMattersBinnedPairing[prob_in_gap][GW200115][prob_mass2_NS]
                & $<0.001$
                & \brokenMatterMattersBinnedPairing[prob_in_gap][GW200115][prob_NSBH]
                \\
          \hline

    \end{tabular}
    \caption{
        \label{tab:evs-broken}
        Events classified in the \broken{} formalism.
        Values are the probabilities that each event's component masses lie in a given region of parameter space, considering uncertainty in the component masses as well as in the inferred values of $\mbreak$.
        If $\mbreak$ is interpreted as the delimiter between NS and BH masses, the probabilities can be understood as labeled in the second row.
        The uncertainties in these probabilities are estimated to be less than $1\%$.
        Only events with $>5\%$ probability of containing a NS according to this classification are included in this table.
    }
\end{table*}

\begin{table*}
  \centering
  \begin{tabular}{c c c c c}
          \hline \hline
          \textbf{Probability}
                & $P( m_2 \leq \gamma_{\text{low}})$ 
                & $P( m_1 \leq \gamma_{\text{low}})$
                & $P(\gamma_{\text{low}} \leq m_2 \leq \gamma_{\text{high}})$ 
                & $P(\gamma_{\text{low}} \leq m_1 \leq \gamma_{\text{high}})$ \\
          \hline
          \textbf{Interpretation}
                & $P(\text{contains a NS})$
                & $P(\text{is a BNS})$
                & $P(m_2 \text{ in mass gap})$
                & $P(m_1 \text{ in mass gap})$ \\
          \hline \hline
          GW170817
                & $>0.99$
                & $>0.99$
                & $<0.001$
                & $<0.001$
                \\
          \hline
          GW190425
                & $>0.99$
                & \dipbreakMatterMattersBinnedPairing[prob_in_gap][S190425z][prob_mass1_NS]
                & \dipbreakMatterMattersBinnedPairing[prob_in_gap][S190425z][prob_mass2_in_gap]
                & \dipbreakMatterMattersBinnedPairing[prob_in_gap][S190425z][prob_mass1_in_gap]
                \\
          \hline
          GW190814
                & \dipbreakMatterMattersBinnedPairing[prob_in_gap][S190814bv][prob_mass2_NS]
                & $<0.001$
                & \dipbreakMatterMattersBinnedPairing[prob_in_gap][S190814bv][prob_mass2_in_gap]
                & $<0.001$
                \\
          \hline
          GW190924\_021846
                & $<0.001$
                & $<0.001$
                & \dipbreakMatterMattersBinnedPairing[prob_in_gap][S190924h][prob_mass2_in_gap]
                & \dipbreakMatterMattersBinnedPairing[prob_in_gap][S190924h][prob_mass1_in_gap]
                \\
          \hline
          GW200105
                & \dipbreakMatterMattersBinnedPairing[prob_in_gap][GW200105][prob_mass2_NS]
                & $<0.001$
                & \dipbreakMatterMattersBinnedPairing[prob_in_gap][GW200105][prob_mass2_in_gap]
                & \dipbreakMatterMattersBinnedPairing[prob_in_gap][GW200105][prob_mass1_in_gap]
                \\
          \hline
          GW200115
                & \dipbreakMatterMattersBinnedPairing[prob_in_gap][GW200115][prob_mass2_NS]
                & $<0.001$
                & \dipbreakMatterMattersBinnedPairing[prob_in_gap][GW200115][prob_mass2_in_gap]
                & \dipbreakMatterMattersBinnedPairing[prob_in_gap][GW200115][prob_mass1_in_gap]
                \\
          \hline
    \end{tabular}
    \caption{
        \label{tab:evs-dipbreak}
        Events classified in the \dipbreak{} formalism.
        Values are the probabilities that each event's component masses lie in a given region of parameter space, considering uncertainty in the component masses as well as in the inferred values of $\glo$ and $\ghi$.
        If $\glo$ and $\ghi$ are interpreted as the lower and upper edges of the mass gap and $\glo$ is therefore interpreted as the maximum observed NS mass in merging binaries, the probabilities can be understood as labeled in the second row.
        The uncertainties in these probabilities are estimated to be less than $1\%$.
        Only events with $>5\%$ probability of being in one of these categories are included in this table.
    }
\end{table*}

\subsection{GW190814}
\label{sec:loo-0814}
GW190814 was observed in the first half of the third observing run \citep{abbott_gw190814_2020} and has the most asymmetric mass ratio of any gravitational wave event to date.
Its secondary has a mass of $m_2 = 2.59^{+0.08}_{-0.09} \Msun$, leaving the nature of this component ambiguous: depending on the choice of maximum allowed NS mass, it can either be classified as a low-mass BH or the most massive NS observed to date.
Previous ``leave-one-out'' studies \citep{abbott_population_2021, essick_probing_2021} have definitively identified GW190814 as an outlier with respect to the detected BBH population.
Additionally, \cite{landry_mass_2021} find that including the secondary of GW190814 in the population of neutron stars in merging binaries considerably changes the inferred mass distribution when compared to a fit that excludes this event, though uncertainties in both fits are large.
It is therefore desirable to find a model that can account for such an extreme event.

\begin{figure*}
    \centering
    \includegraphics[width=1.0\textwidth]{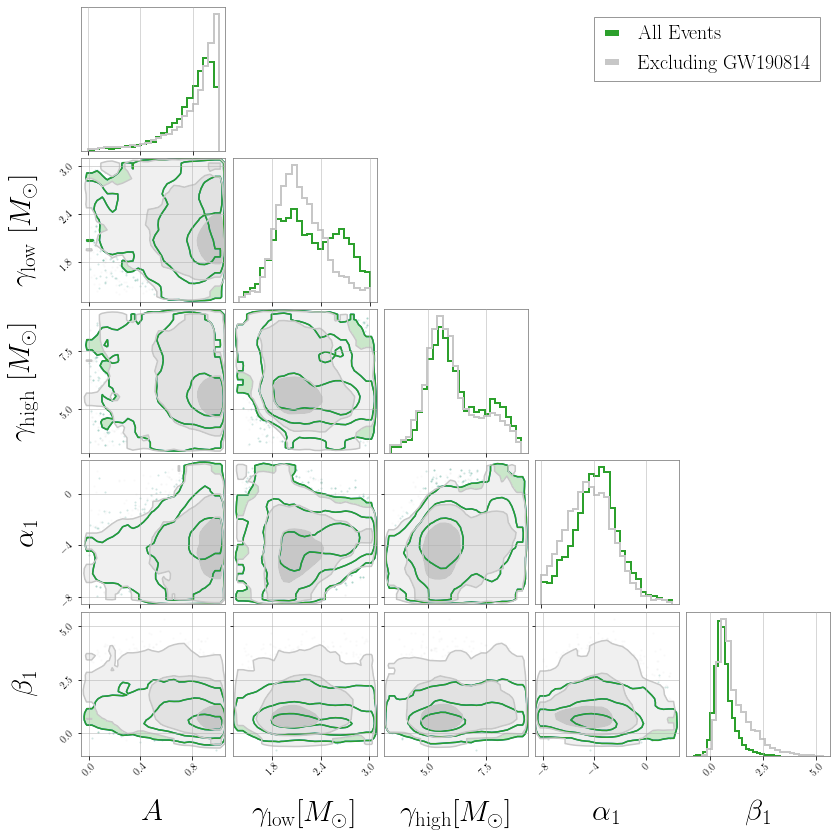}
    \caption{
        Effects of leaving out event GW190814 under the \dipbreak{} model.
        Green contours correspond to the inference when GW190814 is included, and grey contours correspond to the inference when it is excluded.
        Almost all hyperparameters remain unchanged with the inclusion vs exclusion of GW190814, indicating that this event is consistent with the inferred population using this model.
        The only parameter that exhibits a noticeable change is $\glo$, where the bimodality of the posterior distribution disappears with the exclusion of GW190814.}
    \label{fig:loo_0814_mbreak_is_glo}
\end{figure*}

We perform a similar study with GW190814 by comparing the mass distributions inferred with the full event list to those inferred by excluding GW190814.
We find that in the case of both \broken{} and \dipbreak, the inferred distributions without GW190814 are consistent with those with GW190814.
In the framework of \broken, the posterior of $\mbreak$ is largely unchanged with the exclusion of this event, but the posterior distribution of $\beta_1$ shifts to higher values when excluding GW190814.
In the framework of \dipbreak, the posterior of $\glo$ when GW190814 is included in the inference is in some disagreement with that inferred without GW190814.
When GW190814 is included, the posterior on $\glo$ is bimodal, with the lower mode corresponding to putting GW190814's secondary in the gap and the higher mode corresponding to locating the gap at higher masses than GW190814's secondary.
Excluding GW190814 keeps only the lower mode.
Additionally, the posterior on $A$ peaks at a slightly higher value when GW190814 is excluded.
However, the inferred values with and without GW190814 are still consistent with one another, as can be seen in Fig.~\ref{fig:loo_0814_mbreak_is_glo}.
Therefore, both \broken{} and \dipbreak{} successfully account for this event.

Note that such leave-one-out analyses can introduce biases that may lead to the false classification of the left-out event as an outlier unless modifications to the likelihood are used to mitigate this effect \citep{essick_probing_2021}.
However, since these modifications tend to bring the hyperposterior inferred without the event in question closer to that inferred with all events, we assume the conclusions derived from Fig.~\ref{fig:loo_0814_mbreak_is_glo} will remain the same with the use of the methods presented in \citet{essick_probing_2021}. 

\subsection{GW200105 and GW200115}
\label{sec:nsbh}

GW200105 and GW200115 are the most definitive detections of NSBH-like systems via gravitational waves to date, observed by the LVK during the second half of the third observing run \citep{abbott_observation_2021}.
GW200105 has component masses $m_1=8.9_{-1.5}^{+1.2}\, \Msun$ and $m_2 =1.9_{-0.2}^{+0.3}\, \Msun$, and GW200115 has component masses $m_1=5.{7}_{-2.1}^{+1.8}\,\Msun$ and $m_2=1.{5}_{-0.3}^{+0.7}\,\Msun$.
Both events therefore have components near the edges of the purported lower mass gap, and analyzing them in the context of the \dipbreak{} model is helpful in understanding their classification.
Likewise, these events play a large role in probing the values of $\glo$ and $\ghi$ inferred in this analysis.

We first determine if GW200105 and GW200115 are consistent with the rest of the population of CBCs by performing a model fit to \dipbreak{} without these events and comparing it to the population inferred with these events.
Appendix~\ref{sec:loo_nsbh} shows these two fits to the data.
We find that the two inferred populations are consistent with one another, and that the NSBH events have the effect of constraining $\glo$ and $\ghi$ towards a narrower gap.
This effect, as well as the fact that both events have $P(\glo \leq m_i \leq \ghi) < 0.5$, indicate that GW200105 and GW200115 are both consistent with being ``gap-straddling binaries,'' though there is still a definite possibility that either or both events have one component in the lower mass gap.
\\
\\

\section{Discussion}
\label{sec:discussion}

A lower mass gap between NSs and BHs has been suggested by observations of X-ray binaries that find a lack of black holes below $5\, \Msun$ \citep{bailyn_mass_1998,ozel_black_2010, farr_mass_2011} despite predictions for the maximum allowed neutron star mass by the neutron star equation of state \citep{margalit_constraining_2017, ruiz_gw170817_2018, shibata_constraint_2019, chatziioannou_neutron-star_2020, legred_impact_2021} of $\sim 2$ \mdash{} $2.5 \, \Msun$.
Such a dearth of systems just above the maximum neutron star mass is in contrast with theoretical studies of compact object formation scenarios which predict a continuous distribution of supernova remnant masses that have a smooth transition from NSs to BHs \citep[e.g.,][]{fryer_theoretical_2001,fryer_compact_2012, zevin_exploring_2020, drozda_black_2020}.

In this work, we explore whether there is evidence for a lower mass gap in the existing gravitational wave data.
Inferences on the edges of such a gap have previously been carried out by examining the extrema of separate black hole and NS distributions.
However, this approach lessens the amount of information available to both analyses, and can lead to biased inferences. It is insufficient to simply look at the largest neutron star mass and smallest black hole mass in the catalog, and make statements about a mass gap. {\em The best estimates for individual compact object masses can only be achieved through an analysis of the full population, including systems on both sides of the gap as well as those in the gap (if any).}
Fitting a single model across the entire spectrum of compact binary coalescences allows us to find features that are difficult to infer from fits to the BBH, BNS, or NSBH mass distributions alone.

We present global fits to the mass distribution of NSs and BHs in merging binary systems in Fig.~\ref{fig:trace_plots}.
All models considered find a sharp drop in the mass distribution at $\sim 2$\mdash{}$3\, \Msun$, either via a steep power law ({\em top panel}) or a sharper ``discontinuity'' ({\em middle and bottom panels}). 
Although this feature has been determined solely from the mass distribution of GW systems, it presumably relates to the maximum mass of neutron stars in GW binaries. It is suggestive that this break occurs precisely in the mass range predicted from equation-of-state calculations \cite{legred_impact_2021}.
More detections at NS masses will inform the shape of the mass spectrum below $3\, \Msun$, allowing us to better resolve the nature of this drop-off.

For models that allow for a feature in the mass distribution after the initial ``neutron star'' drop, we find a subsequent rise at $\sim 4.5$\mdash{}$8.5\, \Msun$ ({bottom panel of Fig.~\ref{fig:trace_plots}}).
While this hints at a potential mass gap between neutron stars and black holes, we cannot confidently detect nor rule out the existence of such a feature with the limited detections and low sensitivity in that region.

As we emphasize above, understanding a single GW event can only be done in the context of the full underlying population.
Given our population model which bridges the gap between NSs and BHs, we are able to analyze individual events in the context of this model, and thereby classify them as BNS, NSBH, BBH, or binaries with one component in the mass gap.

Using this population-based classification scheme, we find that GW200105 likely straddles the mass gap, with its primary component confidently higher than the inferred upper edge, and the secondary likely falling below the edge, though uncertainties on the lower edge location make the exact classification uncertain.
We similarly reaffirm that GW200115 is an NSBH, with its secondary component below the lower edge of the mass gap and its primary having roughly equal support below and above the upper edge.
Our inference is unable to determine if GW190814's secondary mass lies below or above the lower edge of the mass gap, but slightly prefers it to lie within the mass gap. 
This ambiguity is driven by uncertainty in the location of the lower edge of the gap.
Thus, the nature of GW190814 may emerge as future data constrain global population models. 
We re-emphasize that these results are based solely on GW data; the existence of baryons has not been assumed in this analysis. Nonetheless, the conclusions are broadly consistent with expectations from NS EOS studies as well as analyses of galactic X-ray binaries.

All of the models considered in this work can accommodate the full list of observed events without any obvious outliers.
In particular, we find that GW190814 is consistent with the full set of compact binary coalescences, even if we are not yet able to identify to which subpopulation it belongs.
In this work we focused on the component mass distribution to identify the component objects of binaries. However, future investigations of the mass ratio distribution may help us distinguish between NSBH and BBH population with NSBH systems in recent population synthesis work tending to prefer mass ratios $<0.5$ \citep{broekgaarden_impact_2021}.

The delineation between neutron stars and black holes is a fundamental observable of GW binary populations. 
Observable constraints will inform the astrophysics of compact object formation, neutron star equation-of-state studies, and the formation channels of GW binaries \citep[e.g.][]{fryer_limiting_2002, belczynski_missing_2012, fryer_compact_2012, muller_simple_2016, breivik_constraining_2019, shao_population_2021, liu_final_2021}.
For example, the confident identification of a feature in the overall mass distribution that is distinct from the maximum neutron star mass inferred from nuclear theory would have important implications for specific supernova explosion mechanisms \citep{fryer_compact_2012, kiziltan_neutron_2013}. 
From GW observations alone, we have identified a sharp drop in the mass distribution of compact objects at $\sim2$\mdash{}$3\,\Msun$, as might be expected from the maximum allowed mass of neutron stars. There is insufficient data to confidently constrain the existence of a subsequent mass gap between neutron stars and black holes. 
However, with the second half of LVK's third observing run completed, and a fourth observing run planned, the GW community is positioned to definitively resolve or disprove this mass gap. 


\begin{acknowledgements}
The authors graciously thank Colm Talbot for significant assistance with modifying and running \texttt{gwpopulation},  Mike Zevin for several helpful comments on the manuscript, and 
Daniel Wysocki, Richard O'Shaughnessy, and Will Farr for useful insights on early results.
A.F. is supported by the NSF Research Traineeship program under grant DGE-1735359
M.F. is supported by NASA through NASA Hubble Fellowship grant HST-HF2-51455.001-A awarded by the Space Telescope Science Institute.
R.E. thanks the Canadian Institute for Advanced Research (CIFAR) for support.
Research at Perimeter Institute is supported in part by the Government of Canada through the Department of Innovation, Science and Economic Development Canada and by the Province of Ontario through the Ministry of Colleges and Universities.
D.E.H is supported by NSF grants PHY-2006645 and PHY-2110507, as well as by the Kavli Institute for Cosmological Physics through an endowment from the Kavli Foundation and its founder Fred Kavli.
D.E.H also gratefully acknowledges the Marion and Stuart Rice Award.
This material is based upon work supported by NSF LIGO Laboratory which is a major facility fully funded by the National Science Foundation.
The authors are grateful for computational resources provided by the LIGO Laboratory and supported by National Science Foundation Grants PHY-0757058 and PHY-0823459.
List of software: \texttt{gwpopulation} \citep{talbot_parallelized_2019} \texttt{bilby} \citep{ashton_bilby_2019}, \texttt{numpy} \citep{harris_array_2020}, \texttt{xarray} \citep{hoyer_xarray_2017}, \texttt{matplotlib} \citep{hunter_matplotlib_2007}.

\end{acknowledgements}


\bibliography{references}


\appendix



\section{Most flexible model fit}
\label{sec:all_params_free}
As a check that the parameters we choose to fix are set to values allowed by the data, we perform an analysis where all hyperparameters are simultaneously fit.
Notably, we find $\glo=2.23^{+0.43}_{-0.38}\,\Msun, \ghi=6.34^{+1.18}_{-1.03}\,\Msun,$ and $A=0.89^{+0.08}_{-0.26}$, all of which are broadly consistent with the corresponding hyperposteriors inferred using $\dipbreak$.

A prior range for $\eta_{\{\text{high,low}\}}$ that included both negative values and 50 was too large to allow convergence of our sampler given the relatively low number of detections, so we chose uniform priors between -4 and 12.
Though neither hyperparameter is particularly well-measured, we find both $\eta_{\rm low}$ and $\eta_{\rm high}$ to be positive with $>90\%$ credibility, indicating the existence of a dip in the mass distribution between $\glo$ and $\ghi$: $\eta_{\rm low} = 7.94^{+2.70}_{-3.63}$ and $\eta_{\rm high} = 8.54^{+2.36}_{-3.52}$.
Similarly, we find a Bayes Factor of $\mathcal{B}^{\eta_{\rm high} = 12}_{\eta_{\rm high}=0} = 13$ in favor of a rise in the mass distribution at $\ghi$.
However, both hyperposteriors are approximately uniform after $\sim 8$, indicating that the inference is unable to resolve the steepness of the gap edges beyond a power law that is increasing (or decreasing, in the case of $\eta_{\rm low}$) faster than $\sim m^8$. 
We therefore conclude that fixing the gap edges to be relatively sharp ($\eta_{\{\text{high,low}\}} = 50$) is allowed by the data.

Within this fit, we find that 80\% of draws from the hyperposterior have support for a local minimum between $1 \Msun$ and $10 \Msun$, compared to 49.1\% of prior draws.
This gives a Bayes factor of $1.63$ in favor of a local minimum, showing that a lower mass gap is slightly but not definitively preferred by the data.


\section{Hyperposteriors Inferred With and Without NSBH Events.}
\label{sec:loo_nsbh}

\begin{figure*}
    \centering
    \includegraphics[width=1.0\textwidth]{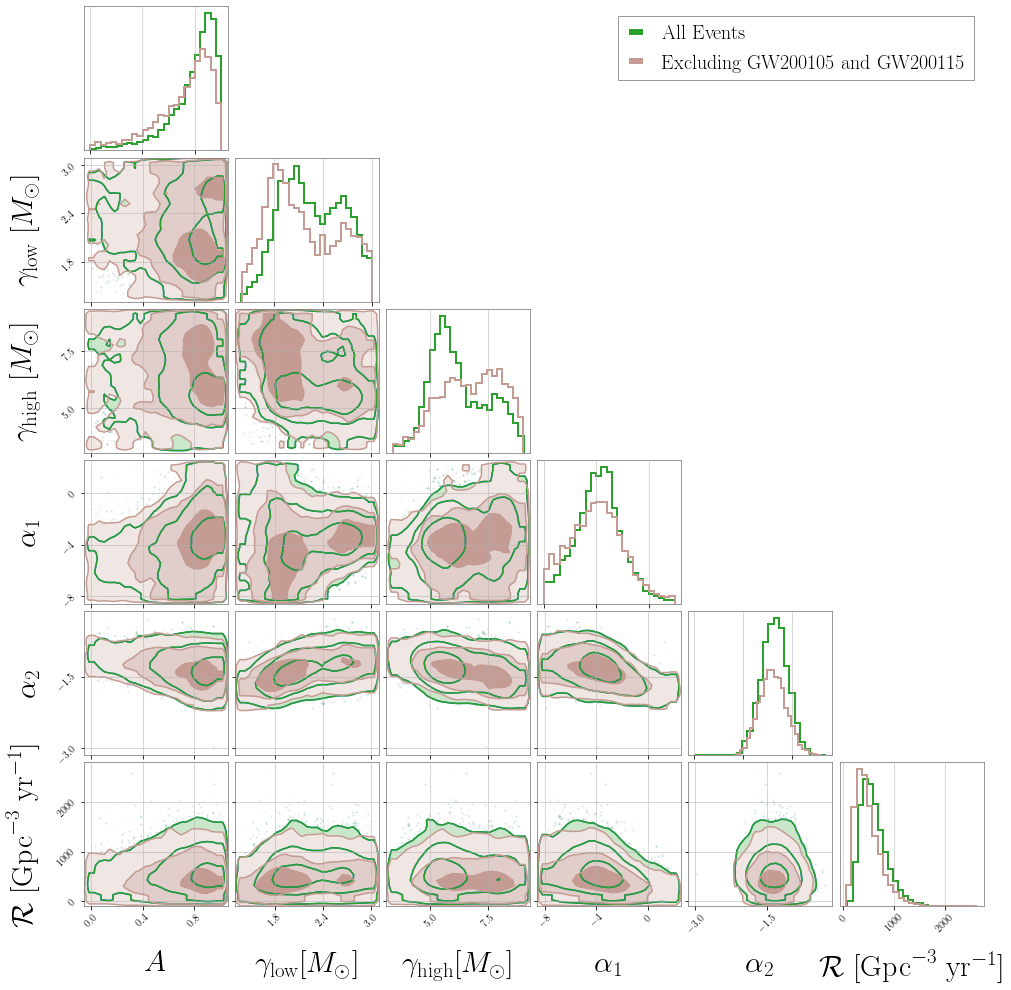}
    \caption{
        \label{fig:loo_nsbh}
        A subset of hyperposteriors for the \dipbreak{} model inferred with (\emph{green}) and without (\emph{brown}) the events GW200105 and GW200115.
        The mass distribution does not change significantly with the exclusion of these events, indicating that the NSBHs are consistent with the rest of the detected population.
        When the NSBHs are included, the upper edge of the mass gap, $\ghi$, is slightly better constrained and support for a wider gap (low $\glo$ and high $\ghi$) diminishes.
        This suggests that these events probe the gap location.
        Hyperparameters on which the exclusion of the NSBH events has no effect are not included in this plot.
        }
\end{figure*}

In Fig.~\ref{fig:loo_nsbh} we present a corner plot of with the posteriors of relevant hyperparameters inferred by including and excluding the NSBH events GW200105 and GW200115. We find these events are broadly consistent with the rest of the CBC population. 
The bottom row shows the effect on the overall rate of mergers.
It illustrates the effects of treating these two events as if they were detected in the first half of the third observing run.
This choice introduces a systematic bias on the rate inference of $23\%$, which is well within the statistical uncertainties of the rate.
Nonetheless, we do not quote merger rates of NSBH-like events, leaving this calculation to future analyses that are able to include the full set of GW detections in the third observing run.

\end{document}